\documentclass[pra,superscriptaddress,twocolumn]{revtex4}

\usepackage[dvips]{graphicx}%
\usepackage{amsmath}
\usepackage{bm}
\begin{document}
%%%%%%%%
%  \small
%%%%%%%%

\title{Measurement schemes for the spin quadratures on an ensemble of atoms}

%\author{Ryo Namiki}% / \today
%\email[Electric address: ]{namiki@qi.mp.es.osaka-u.ac.jp}%{namiki@qo.phys.gakushuin.ac.jp} 
%
%\author{Tetsushi Takano}\author{Yoshiro Takahashi} %\email{takano@scphys.kyoto-u.ac.jp}
%\affiliation{Department of Physics, Graduate School of Science, Kyoto University, Kyoto 606-8502, Japan}
%\author{Shin-Ichi-Ro Tanaka}
%\affiliation{Department of Physics, Graduate School of Science, Kyoto University, Kyoto 606-8502, Japan}
\author{Ryo Namiki}\affiliation{Department of Physics, Graduate School of Science, Kyoto University, Kyoto 606-8502, Japan}
\author{Tetsushi Takano}\affiliation{Department of Physics, Graduate School of Science, Kyoto University, Kyoto 606-8502, Japan}
%\affiliation{Photon Science Center, School of Engineering, The University, of Tokyo, 7-3-1 Hongo, Bunkyo-ku, Tokyo 113-8656, Japan}
\author{Shin-Ichi-Ro Tanaka}\affiliation{Department of Physics, Graduate School of Science, Kyoto University, Kyoto 606-8502, Japan}
\author{Yoshiro Takahashi}\affiliation{Department of Physics, Graduate School of Science, Kyoto University, Kyoto 606-8502, Japan}\affiliation{CREST, JST, 4-1-8 Honcho Kawaguchi, Saitama 332-0012, Japan}

\date{May 8,2009}%\today}%March 17, 2007}%
\begin{abstract} 
We consider how to measure collective spin states of an atomic ensemble based on the recent multi-pass approaches for quantum interface between light and atoms. We find that a scheme with two passages of a light pulse through the atomic ensemble is efficient to implement the homodyne tomography of the spin state. Thereby, we propose to utilize optical pulses as a phase-shifter that rotates the quadrature of the spins. This method substantially simplifies the geometry of experimental schemes.
\end{abstract}

%\pacs{03.67.Dd, 42.50.Lc} 
%\keywords{ }
\maketitle

%\newpage
%%%%%%%%%%%%%%%%%%%%%   part 1 %%%%%%%%%%%%%%%%%%%%%%%%%%%%%%%%%%%%%%%%%%%%%%%%
%%%%%%%%%%%%%%%%%%

%\small 

%Abstract: \textit{}
% \section{Introduction}
To establish a method for determining the state of a quantum system is an important step for progress in modern physics. For light fields on a few well-defined modes, the quantum-state tomography to determine the density matrix of the system has been developed mainly based on the quadrature measurement implemented by a balanced homodyne detection \cite{RMP-tomography,leonhardt}. The quadrature measurement is a standard tool in the approaches for continuous variable quantum information  \cite{RMP-CV}. However, outside of the optical systems it seems difficult to tell how to implement such a measurement.

It has been known that a continuous variable description of the system can be applied on the collective-spin state of an ensemble of massive particles, and there exist various approaches for implementing continuous variable quantum information processing with ensembles of atoms \cite{Bra-Pati,Cerf-Leuch-Polzik}. In particular, optical accessibility and coherent properties of the atomic ensembles are thought to be useful for implementing quantum memory of light and constructing a quantum interface to transfer the quantum state between the light and atoms \cite{Bra-Pati,memoryExp,tele,EPR,retr,Kuzmi,Takano}. %

In order to construct a quantum interface, one of the central atom-light interaction is the Faraday-Rotation (FR) interaction \cite{QND,Kuzmi}. %(which is also referred to as a CV NOT gate). 
  On one hand, this interaction provides a two-mode-coupling gate to essentially implement any quadratic Hamiltonian interaction \cite{Fi03,Kurotani-Ueda}.
   While the design of the interaction Hamiltonian and achievable fidelities in the gate operations of interface are widely investigated as well as generation schemes of quantum entanglement \cite{Cerf-Leuch-Polzik,Mus}, how to estimate such a gate operations experimentally is less concerned \cite{Rig06,Ham05,namiki07,Takano}. One of the reasons for this might be the lack of an established measurement scheme to probe the spin state such as the homodyne measurement in optics. The measurement of the spin state and state reconstruction to visualize its quantum nature in the form of the quasi-probability distribution are interesting objective in its own right \cite{Usami06}.
 On the other hand, the FR interaction corresponds to a unitary operation used in a classical model for an indirect measurements \cite{Neumann}, and recently several models of quantum measurements have been proposed to demonstrate the relation between an indirect measurement and its back action \cite{Ozawa88,Ozawa03,Kitano08}.

 Until recently the construction of the quantum interface based on the multiple use of the FR interaction \cite{Bra-Pati,retr,Takano} has been thought to be impractical due to the requirement of a long delayline (DL) to storage the pulse until an interaction is completed.  %\ 
However, an experimental demonstration of the FR interaction using a short pulsed light of a few hundred-nano-second width and an ensemble of Laser-cooled atoms has been reported  \cite{Yb-QND08}, % (who has a long lived spin-one-half ground state of Nuclear spin)  
and thus the required length of the DL is thought to be a feasible size. %(M.)
In this report we consider measurement schemes for the tomographic reconstruction of collective spin states and make a link to the measurement theory.

\begin{figure}[tb]
  \begin{center}
    \includegraphics[width=7.5cm]{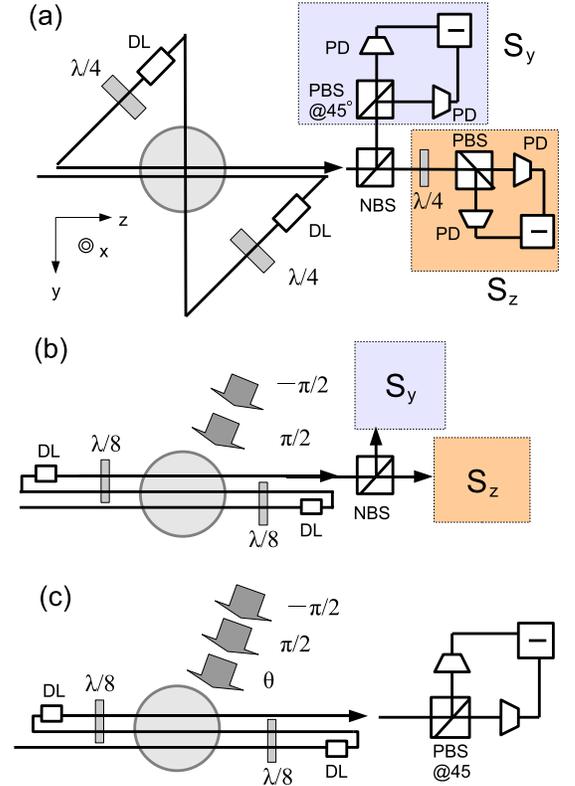}%{opt1.eps}
  \end{center}
  \caption{The interaction induced by the three passages of the light through the atomic ensemble exchanges the quadrature states of the light and atomic collective spins. (a) The Husimi-Q function of the outgoing light is measured by a double homodyne measurement. The delaylines (DL) are required to storage the light pulse in order to eliminate the overlaps of the interactions. %
   The diagonal polarization $S_y$ of one half of the pulse split by a non-polarized beamsplitter (NBS) is measured by a 45-degree inclined polarized beamsplitter (PBS) and differentiation of the two photo detector's (PD) outputs.  The circularly polarization $S_z$ of the other half of the pulse is measured by the other homodyne detector with a quarter-wave plate $\lambda /4$. (b) Folding-path setup for the measurement of the Husimi-Q function with the help of $\pm\pi/2$-spin rotations via fictitious magnetic fields. (c) Folding-path setup for the homodyne tomography of the spin state. A phase rotation $\theta$ is applied on the spins before the interaction in order to measure the quadrature of various angle.}%
    \label{fig:Fig1.eps}
\end{figure}

%\subsubsection{Stokes and quadrature measurement via polarization difference}
Let us write the Stokes operator of a pulsed light propagating along with $z$ axis   \begin{eqnarray}
S_x =   (\hat a_H^\dagger \hat a_H -\hat a_V^\dagger \hat a_V)/2 \nonumber \\
S_y =  (\hat a_D^\dagger \hat a_D -\hat a_{\bar D}^\dagger \hat a_{\bar D})/2 \nonumber \\
S_z =  (\hat a_+^\dagger \hat a_+ -\hat a_-^\dagger \hat a_-)/2,
\end{eqnarray} where $\hat a_{H(V)}$ is the annihilation operator of the pulse mode with horizontally (vertically) polarization, which is related to the operators for the diagonally polarization  $\hat a_D = (\hat a_H + \hat a_V) /\sqrt 2  $, $\hat a_{\bar D} = (\hat a_H - \hat a_V) /\sqrt 2  $, and the circularly polarization $\hat a_\pm  = (\hat a_H \pm i \hat  a_V) /\sqrt 2  $. % the operators for the polarized component with res diagonally/circular are given by .
The Stokes operators satisfy the angular momentum commutation relation, $[S_x, S_y]=i S_z$. % is the and the components of spin operator $J$ of atomic spins 
If the horizontally polarized light is considered to be a strong local oscillator (LO) field with almost fixed average photon number $n_H =\langle \hat a_H^\dagger \hat a_H  \rangle \gg 1$ and phase $\phi$, i.e. $\hat a_H \simeq \sqrt {n_H} e^{-i\phi}$, we can write the difference of the diagonally polarized photon numbers as
\begin{eqnarray}
S_y     &=& (\hat a_V^\dagger \hat a_H +\hat a_H^\dagger \hat a_V)/2 \nonumber \\ 
        &\simeq & \frac{\sqrt{ n_H} }{ 2}(\hat a_V e^{-i\phi} +\hat a_V^\dagger  e^{i\phi}  )\nonumber \\
       &= & \sqrt \frac{ n_H}{ 2}(\hat x_V \cos {\phi} -\hat p_V \sin {\phi}  ) \label{sy}       , 
\end{eqnarray} where we defined the quadrature operators of the vertical mode, 
%\begin{eqnarray}
$\hat x_V  \equiv   {(\hat a_V +\hat a_V^\dagger  ) }/{\sqrt 2}$ and
$\hat p_V  \equiv   {(\hat a_V - \hat a_V^\dagger  ) }/{\sqrt 2 i} $.
%\end{eqnarray}
Similarly, the difference of the circularly polarized photon numbers becomes 
\begin{eqnarray}
S_z     &=& -i (\hat a_V^\dagger \hat a_H -\hat a_H^\dagger \hat a_V)/2 \nonumber \\ 
& \simeq & - \sqrt \frac{ n_H }{ 2}(\hat x_V \sin {\phi} +\hat p_V \cos {\phi}  ) \label{sz}.
 \end{eqnarray}
From Eqs. (\ref{sy}) and (\ref{sz}), one can see that the normalized Stokes operators $ {s_y} = S_y/\sqrt{n_H /2 },  {s_z}  = S_z/\sqrt{n_H /2}$ play the role of the canonical continuous variables, $[s_y, s_z ] =i $. The measurement of the polarization difference gives the statistics of the variable, and it acts as the quadrature measurement of quantum optical mode [see FIG. \ref{fig:Fig1.eps} (a)]. 
%\

In the optical homodyne tomography, the phase difference between the signal and LO fields $\phi$ is modulated by the amount of $\delta \in [0, \pi]$.
However, it might be difficult to separate the two polarization components and/or apply the phase shift on each of the polarization modes with sufficient resolution individually when the intensities of the two polarizations are highly different so that $n_V=\langle \hat a_V^\dagger \hat a_V \rangle  \ll n_H  $ \cite{hirano03}.
For such a collinear propagation of the light, the double homodyne measurement might be effective. In this measurement the $S_z$ measurement and $S_y$ measurement are performed on the pulses equally split by the non-polarized beamsplitter as in FIG. \ref{fig:Fig1.eps} (a). It is known that the double homodyne measurement gives the Husimi-Q function and is tomographic complete \cite{leonhardt}. In order to see this, let us write the density operator of the vertical light with the P representation, $\hat \rho = \int \textrm{d}^2\alpha P(\alpha )|\alpha \rangle \langle \alpha |$ where $| \alpha \rangle $ represents the coherent state. The $x$-quadrature and $p$-quadrature  distributions of $|\alpha \rangle$ are given by $|\langle x |\alpha \rangle |^2  = \exp\{-(x- \sqrt 2 \textrm{Re} [\alpha ])^2 \}/\sqrt \pi $ and $| \langle p  |\alpha  \rangle  |^2   = \exp\{-(p- \sqrt 2 \textrm{Im} [\alpha ] )^2\}/\sqrt \pi $, respectively. Due to the beamsplitter transformation $|\alpha \rangle |0\rangle_{\textrm{vac}}  \to |\alpha /\sqrt 2\rangle  |\alpha /\sqrt 2 \rangle_{\textrm{vac}} $, the double homodyne measurement that measures $x$ quadrature and $p$ quadrature on each of the modes yields the probability distribution $Q(x,p) \equiv  \int \textrm{d}^2 \alpha P(\alpha )  |\langle x |\alpha /\sqrt 2\rangle |^2  | \langle p  |\alpha /\sqrt 2 \rangle_{\textrm{vac}}  |^2 =\int \textrm{d} ^2\alpha P(\alpha ) \exp (-|x+ip - \alpha |^2) /\pi  $. This convolution of the P function corresponds to the Husimi-Q function since $\langle \alpha |\hat \rho | \alpha \rangle / \pi = \langle \alpha |\int  P(\alpha ')|\alpha' \rangle \langle \alpha' |\textrm{d}^2\alpha'|  \alpha \rangle / \pi= Q(\textrm{Re} [\alpha ], \textrm{Im}[ \alpha ] )$. %= \int  P(\alpha ') \exp(|\alpha -\alpha'|^2 ) d^2 \alpha '$.  %$Q(x,p) = \langle \alpha | \hat \rho |  \alpha  \rangle \big |_ {\alpha =(x+ip)/\sqrt2} $,
 Therefore, one can directly reconstruct the quantum state in the form of the Husimi-Q function. % 
 
%\subsubsection{Spin operator}

Let us write the collective spin operator composed of an ensemble of spin-one-half particles defined by the population difference between the spin-up state $|\uparrow\rangle$ and spin-down state $|\downarrow \rangle$ quantized in $x$ direction as   
 \begin{eqnarray}
J_x &=&   (|\uparrow\rangle\langle\uparrow | -| \downarrow\rangle\langle \downarrow| )/2 \nonumber \\
J_y &=&  (| \uparrow \rangle\langle  \downarrow|  +|\downarrow \rangle \langle \uparrow| )/2 \nonumber \\
J_z &=&  i(| \uparrow \rangle\langle  \downarrow|  -|\downarrow \rangle \langle \uparrow| )/2.  
\end{eqnarray} %The angular momentum commutation relation $J \times J = i J$ holds. 
 We assume the number of the particles $J = 2 \left\langle  J_y^2 +J_z^2\right\rangle \gg 1$ and the spins are almost polarized in $x$ direction, e.g. $\langle J_x \rangle  \sim   \frac{J}{2}$.  Noting that the angular-momentum commutation relation holds $[J_y,J_z] =i J_x$  %due to the commutation relation of annihilation and creation operators. Similar to the case of the Stokes operator,
 the normalized spin operators, $ {j_y} = J_y/\sqrt{N/2}$ and ${j_z} = J_x/\sqrt{J}$, act as the canonical continuous variable, $[j_y, j_z] =i $,  similar to the case of the Stokes operator. We refers to $j_y$ and $j_z$ as the \textit{spin quadratures}.% 

%\subsubsection{FR interaction and Multi-pass approach for Light-Atom interface}
The action of the FR interaction due to the passage of the light through the ensemble %whose collective spins $J := (J_x, J_y, J_z) $ 
is described by the unitary operator
\begin{equation}
U(\alpha t) =\exp (-i\alpha t S_z J_z ) \label{fr1},
\end{equation}
where $\alpha$ is a coupling constant and $t$ is an interaction time. %the interaction strength $\kappa $ is given by $\kappa =  \alpha t \sqrt{SJ}$.
This FR interaction rotates both of the light and spin quadratures about the light propagating axis ($z$ axis). For the quantum interface, the parameters are selected to attain $\kappa \equiv  \alpha t \sqrt{SJ} = 1 $. % is introduced the other parameters are chosen to achieve $\kappa \sim 1$.
 With the normalized operators we concerns the elementary gate interaction 
\begin{equation}
U_z =\exp (-i s_z j_z ).  \label{time2} 
\end{equation}

In the recent proposal \cite{Takano}  (see also \cite{Kurotani-Ueda,Fi03}), a swapping gate that exchanges $(s_y,s_z)$ and $(j_y,j_z)$ can be implemented by the three passages of the light through the atomic-spin ensemble as in FIG. \ref{fig:Fig1.eps} (a). %
Obviously, if the swapping is performed, the measurement of the light quadrature implies the measurement of the quadrature of the initial spin state. Hence,
 after the swapping operation, one can apply the quantum optical measurement on the light in order to determine the density operator of the initial spin state. 
 Then an application of the double homodyne measurement as in FIG. (a) enables to reconstruct the spin state in the form of the Husimi-Q function directly. 
   
 This measurement might be preferable to show the better-than-classical performance of the transfer/storage gate with the input of the coherent states \cite{Takano,Ham05,namiki07}. The measured probability distribution $\langle \alpha | E(\hat \rho_{\textrm{in}} )|    \alpha \rangle$ enables us to calculate the average fidelity $\bar F(E,\eta, \lambda)\equiv  \int d^2 \alpha p_\lambda( \alpha ) \langle \sqrt \eta\alpha| E \left( |\alpha \rangle\langle \alpha|  \right) |\sqrt \eta \alpha \rangle $, where $E$ represents the gate action and $p_\lambda( \alpha ) =  \frac{\lambda }{\pi} \exp (- \lambda |\alpha |^2 )$ is a Gaussian probability  distribution. % introduced to show the non-classical performance within the sampling of finite-amplitude input states. % on the input state $\hat \rho_{\textrm{in}}$, $\hat \rho_{\textrm{out}}= E(\hat \rho_{\textrm{in}} )$.
%It is shown that the classical limit fidelity for any $\lambda >0 $ and $\eta > 0$ is given by $F_c({ \eta, \lambda}) \equiv \frac{1+\lambda}{1+\eta +\lambda }$.
 If one can find a pair of positive numbers  $(\lambda,\eta)$ s.t., $\bar F(E,\eta, \lambda) > F_c({ \eta, \lambda}) = \frac{1+\lambda}{1+\eta +\lambda } $ then the gate is capable of transmitting quantum correlation since the gate cannot be simulated by any measure-and-prepare scheme. While $\lambda$ is introduced to test the better-than-classical performance within a finite input state distribution, $\eta$ is introduced to take into account the non-unit-gain (non-unitary) effect of the experiments. %T
It is shown \cite{namiki07} that the independent measurements of the quadratures, $S_y$ and $S_z$, are also useful to check the  criterion since a lower bound of the fidelity to a coherent state can be estimated from the expectation values and variances of the two quadratures. In this case, we may calculate the average quadrature mean-square deviation $ \bar \delta (E,\eta, \lambda)= \int p_\lambda (\alpha )\textrm{Tr}   \{ E\left( |\alpha \rangle\langle \alpha |\right)  [ (  s_y - \sqrt{ 2\eta} \textrm{Re} [\alpha e^{i\phi}])^2+( s_z - \sqrt{2 \eta} \textrm{Im} [\alpha^*e^{-i\phi} ])^2  -1 ] \} \textrm{d} ^2 \alpha $. The inequality  
%For any entanglement breaking channel $E$,
$ \bar \delta (E, \eta, \lambda) <  \frac{2 \eta }{ 1+ \lambda + \eta}$ is a sufficient condition for the non-classical performance of the gate. %\end{eqnarray} 2 (1- F_c(  \eta, \lambda) ) =

 As we have mentioned, the optical homodyne tomography for collinearly propagating light may not be feasible due to the difficulty of the phase rotation. For the tomography of the spin state % Recalling that the quantum state we concern here is originally the state of the spin quadrature, 
 this is not matter if one can apply the phase rotation on the spin state before the swapping operation. % is useful to perform the homodyne tomography. 
 For the implementation of the spin rotation, we propose to apply the FR interaction of Eq. (\ref{fr1}) with a circularly polarized pulse in the strongly polarized direction ($x$ direction). We refers to this phase rotation as the fictitious-magnetic-field phase-shifter (FMPS) because its action is the same as the Larmor precession by a static magnetic field \cite{CCT72,FM}. Due to the optical means, the FMPS has a good local addressability and its action can be quicker than the magnetic-resonance rotation induced by the RF field. Note that the rotation angle can be a sizable amount whereas the rotation angle due to the interaction of Eq. (\ref{time2}) is considered to be small. % 

This FMPS enables us to substantially change the geometry of the experimental setup and will be helpful to manage a limited optical access of the atomic-spin ensembles possibly in a vacuum chamber. Figure \ref{fig:Fig1.eps}(b) illustrates a setup of swapping gate with the FMPS. In this scheme, the second passage of the light in $y$ direction in FIG. \ref{fig:Fig1.eps}(a) can be replaced with the counter propagation along with $z$ axis provided that the spins are $\pi /2$ rotated about $x$ axis after the first passage. Then additional $- \pi /2$ rotation and further counter propagation complete the swapping gate.  The counter propagation scheme is favorable to earn the mode coupling as well as to manage the limited optical access. %
The single quadrature measurement of any spin-quadrature angle can be executed by initially applying the phase rotation $\theta$ onto the spins via the FMPS as schematically shown in FIG. \ref{fig:Fig1.eps} (c). This provides a possible scheme for the homodyne tomography of the spin state.

\begin{figure}[htb]
  \begin{center}
    \includegraphics[width=6cm]{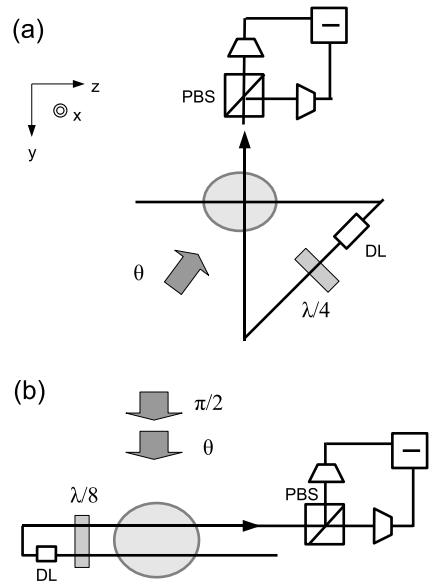}%{opt2.eps}
  \end{center}
  \caption{(a) The two passages of the light pulse through the atomic ensemble exchange a single quadrature of the light with a single quadrature of spin. The swapped single quadrature is measured by a homodyne detector. This measurement with various phase rotation $\theta$ enables to perform the homodyne tomography of the spin state. (b) Folding  setup for the homodyne tomography of the spin state with the help of $\pi/2$ rotation. }% 
  \label{fig:Fig2.eps}
\end{figure}

Although the use of the swapping gate  is a straightforward approach, to demonstrate the three passages of light with two DLs might be difficult.
It is known that the two-pass scheme depicted in FIG. \ref{fig:Fig2.eps} partially exchanges the quadratures \cite{Bra-Pati,Fi03,Kurotani-Ueda,Takano}. The first passage in $z$ direction and the second passage in $y$ direction of the light transform the pair of the quadratures as $(s_{y'}, s_{z'}, j_{y'}, j_{z'})=U_z^\dagger (s_{y }, s_{z }, j_{y }, j_{z }) U_z \simeq(s_{y}+j_z, s_{z }, j_{y }+s_z, j_{z })  $ and  $(s_{y''}, s_{z''}, j_{y''}, j_{z''})=U_{y'}^\dagger (s_{y'}, s_{z'}, j_{y'}, j_{z'}) U_z \simeq(s_{y}+j_z, -j_{y }, j_{y }+s_z, -s_{y })  $, respectively. 
%$S_{z''}  =  -J_y $, $S_{y''}  =   S_y + J_z  $, $J_{z''}  = -S_y $,  and $J_{y''}  =  J_y +S_z$ 
% \begin{eqnarray}
%S_z'' &=& -J_y \\ % (1- \kappa )S_z -\kappa J_y \sim 
%S_y'' &=&  S_y + J_z  \label{sypp} \\ %S_y +\kappa J_z \sim  
%J_z'' &=&-S_y \\ %(1- \kappa )J_z -\kappa S_y \sim 
%J_y'' &=& J_y +S_z %J_y +\kappa S_z \sim  
%\end{eqnarray}
 % and the transfer fidelity tends to unit when the light is squeezed. 
Then the quadrature measurement of $s_{z''}$ is efficient to determine the quadrature distribution of the initial spin, $j_y$.  % Since the homodyne
To be concrete, let us write the Wigner function of the initial system $W_J(j_z,j_y) W_S(s_z,s_y)$ where the subscripts $J$ and $S$ denote the spin and light, respectively. The partial swapping transforms this function as  $W_J(j_z,j_y) W_S(s_z,s_y) \to W''= W_J(j_{z''}+s_{y''},- s_{z''}  ) W_S(  s_{z''}+j_{y''} ,-j_{z''}) $.
Then the measurement statistics of $z$-component of the light quadrature is given by $ \int W'' ds_{y''}dj_{z''} dj_{y''} =  \int  W_J(s_{y''},-s_{z''})  ds_{y''}$. % provided the initial distribution is localized within a finite space:
 This corresponds to the quadrature distribution of $j_y$. % up to the sign of the $j_y$ axis.
  Hence, the partial swapping together with the phase shifter enables us to perform the homodyne tomography to reconstruct the spin state. The two-pass interaction is referred to as the noiseless quadrature transducer \cite{Ozawa03} associated with the contractive state measurement \cite{Ozawa88,Kurotani-Ueda,Kitano08}.
If we fold this two-pass scheme exploiting the FMPS then we have a simpler scheme for the homodyne tomography of the spin state with one DL and two interactions with a counter propagation as in FIG. \ref{fig:Fig2.eps} (b). %

The operations of all above measurement schemes are independent of the initial input state of the light field. An improvement of the measurement accuracy by using squeezed states is one of the central interest in the field of quantum optics \cite{QND,Sor98,Bra-Pati}. In the two-pass scheme, it is known that a better state transfer can be performed if the input light is a squeezed state \cite{Bra-Pati}. % 
Similarly, if we can prepare a well-squeezed light pulse, a better measurement of the spin quadrature is possible with a single passage of the light \cite{QND,Sor98}. For instance, the light quadrature of  the first passage $s_{y'}  \simeq s_y +j_z  $ implies that the measurement of $s_{y'}$ yields the quadrature distribution of $j_z$  provided that the initial light state is infinite squeezed vacuum, s.t., $\Delta s_y \to 0 $  and  $\langle s_y\rangle $ =0. This measurement corresponds to the indirect measurement introduced by von Neumann  \cite{Neumann,Kurotani-Ueda}. Detailed description of measurement schemes associated with a single passage can be found in \cite{Usami06,tele,memoryExp}.% 

In conclusion, we have considered how to measure the quantum state of atomic collective spins based on the multi-pass approach for quantum interface and quantum optical homodyne measurements. % 
Thereby, we have proposed to utilize the FR interaction as a phase shifter for the homodyne tomography, and presented possible modifications for the measurement schemes. 
%W
 We are hoping that those measurement schemes naturally form a part of the multi-pass approach for quantum interface and provide an established tool for the homodyne tomography of the spin state as well as for the test of the quantum measurement theory.

%\begin{theacknowledgments}
This work was supported by the Grant-in-Aid for the Global COE Program ``The Next Generation of Physics, Spun from Universality and Emergence'' from the Ministry of Education, Culture, Sports, Science and Technology (MEXT) of Japan.
 R.N. and T.T. are supported by JSPS Research Fellowships for Young Scientists. 
%\end{theacknowledgments}

\end{document}